\documentstyle[aps,prl,epsf,floats]{revtex}

\draft

\begin{document}
\twocolumn[\hsize\textwidth\columnwidth\hsize\csname
@twocolumnfalse\endcsname

\title{Strong constraint on large extra dimensions from cosmology}

\author{Steen Hannestad}

\address{NORDITA, Blegdamsvej 17, DK-2100 Copenhagen, Denmark}

\date{\today}

\maketitle

\begin{abstract}
We have studied cosmological constraints on the
number and radii of possible large extra dimensions.
If such dimensions exist, 
Kaluza-Klein (KK)
modes are copiously produced at high temperatures in the early
universe,
and can potentially lead to unacceptable
cosmological effects. We show that during reheating, 
large numbers of KK modes are produced. 
These modes are not diluted completely by the entropy production
during reheating because they are produced 
non-relativistically. 
This means that the modes produced during reheating can easily
be the dominant component.
For instance, for two extra dimensions the bound on their
radii from considering only the thermally produced KK modes
is $R \leq 1.1 \times 10^{-4}$ mm. If the modes produced during
reheating are also accounted for, the bound is strengthened 
to $R \leq 2.2 \times 10^{-5}$ mm. This bound is stronger than
all other known astrophysical or laboratory limits.
\end{abstract}

 \vskip1.9pc]


\section{Introduction}

In the past few years there has been an 
enormous interest in the possibility that the presence of large 
extra dimensions can explain the hierarchy problem
\cite{add98,Antoniadis,add99,hlz99,grw99}, 
the fact that
the energy scale for gravitation (the Planck scale $\sim 10^{19}$ GeV)
 is so much
larger than that for the standard model (100 GeV).
The idea is that the standard model fields are located on a 3+1 dimensional
brane embedded in a higher dimensional bulk, where only gravity is allowed to
propagate. 

This already puts stringent constraints on the size of the
extra dimensions. Newtons law should definitely hold for any scale
which has so far been observed. At present the best experiments
have probed scales down to about 1mm.
Thus, if there
are extra dimensions, they can only appear at a scale smaller than that.
For simplicity we make the assumption that the $n$ new
dimensions form an $n$-torus of the same radius $R_n$ in each
direction
\footnote{This assumption has been made in practically all works
on the subject, however see Ref.\ \cite{kaloper2} for a different
model.}.
If there are such extra dimensions, the Planck scale of the full
higher dimensional space, $M_{P,n+4}$, 
can be related to the normal Planck scale, $M_{P,4}$,
by use of Gauss' law \cite{add98}
\begin{equation}
M_{P,4}^2 = R^n M_{P,n+4}^{n+2},
\end{equation}
and if $R$ is large then $M_{P,n+4}$ can be much smaller than $M_P$.
If this scenario is to solve the hierarchy problem then $M_{P,n+4}$
must be close to the electroweak scale ($M_{P,n+4} \lesssim 
10-100$ TeV), otherwise the hierarchy problem reappears.
This already excludes $n=1$, because $M_{P,n+4} \simeq 100$ TeV 
corresponds 
to $R \simeq 10^{8}$ cm. 
However,
$n \geq 2$ is still possible, and particularly for $n=2$ there is the
intriguing perspective that the extra dimensions could be accessible
to experiments probing gravity at scales smaller than 1 mm.
From this point on we use $M$ instead of $M_{P,n+4}$
to simplify notation.

So far, the strongest constraints come from the observation of 
the neutrino emission of SN1987A \cite{SN1987A,hanhart,hanhart2}. 
In the standard model, a Type II
supernova emits energy almost solely in the form of neutrinos.
Furthermore, the observed neutrino signal fits very well with the 
theoretical prediction. If extra dimensions are present, then
the usual 4D graviton is complemented by a tower of Kaluza-Klein states,
corresponding to the new available 
phase space in the bulk. Emission of these KK states can potentially
cool the proto-neutron star too fast to be compatible with 
observations.
This has lead to the tight bound that
$R \lesssim 0.66 \mu$m 
($M \gtrsim 31$ TeV) for $n=2$ and $R \lesssim 0.8$ nm
($M \gtrsim 2.75$ TeV) for $n=3$ \cite{hanhart2}.
In fact, an even stronger constraint can be obtained from considering
the contribution to the diffuse gamma background from decays of
the KK modes produced in cosmological supernovae. A conservative
estimate yields a bound of 
$R \lesssim 0.09 \mu$m 
($M \gtrsim 84$ TeV) for $n=2$ and $R \lesssim 0.19$ nm
($M \gtrsim 7$ TeV) for $n=3$ \cite{hr01}.

The other obvious place in astrophysics to look for these extra
dimensions is cosmology
\cite{add99,hs99,fair} (see also \cite{afp00}).
In the present paper we go 
through the possible cosmological effects from the presence of
large extra dimensions.
We solve the Boltzmann equation for the production of KK modes,
both during the radiation dominated epoch and during the reheating
phase preceding it.
We show that unless the maximum temperature reached during
reheating is very low, the constraints from cosmology are much
stronger than the supernova bounds.


\section{Boltzmann equations}

The fundamental equation governing
the evolution of all species in the expanding universe is the 
Boltzmann equation \cite{kolb},
$L[f] = C[f]$,
where $L = \partial f/\partial t - p H \partial f/\partial p$ 
is the Liouville operator and $C$ is the collision
operator describing all possible interactions.
$f$ is the distribution function for the given particle species.
In the present case, there are two terms contributing to the
collision operator: production and decay. There are several
possible production channels \cite{add99}: 
gravi-Compton scattering,
pair annihilation and bremsstrahlung. In a supernova, nucleon-nucleon
bremsstrahlung is by far the dominant mechanism because of the
very high nucleon density. 
However, this is not the case in the early universe, 
the reason being that the early universe is a high entropy
environment
($\eta = n_B/n_\gamma \simeq 10^{-10}$) \cite{peacock}. Therefore
NN bremsstrahlung is suppressed by a large numerical factor
$\simeq n_N^2/n_\gamma^2$.
The dominant processes are instead the pair annihilation reactions
$2 \gamma \to KK, \nu \bar\nu \to KK, e^+ e^- \to KK$ \cite{hlz99,hs99}.
The matrix element for each of these processes is given simply by
$\sum |M|^2 = A_i s^2/4 \bar{M}_P^2$,
where $A_\nu = A_e = 1$ and $A_\gamma = 4$ \cite{hlz99,hs99}.
$\bar{M}_P = 2.4 \times 10^{18}$ GeV is the reduced Planck mass.

In order for the standard cosmological equations to apply, it is
a necessary condition that $\rho_{KK} \ll \rho_i$, where $i$
denotes fields on the brane.
This means that we can completely neglect inverse processes
in our treatment. It also has the big advantage that we can
neglect Pauli blocking and stimulated emission factors
in the Boltzmann equation. In this
case we can use the integrated
Boltzmann equation which is much simpler than the full Boltzmann
equation \cite{kolb}
\begin{equation}
\dot{n}_m = \sum_{i=\nu,e,\gamma} \langle \sigma v \rangle_i n_i^2 - 
3 H n_m - \Gamma_{{\rm decay},m},
\end{equation}
where $m$ is the mass of the KK state.
For the relatively low mass modes we look at, the decay lifetime
is very long \cite{hlz99}. Therefore decays can be completely neglected
at early times and the production phase can be separated
from the decay phase.
The production equation is then given by \cite{hs99}
\begin{equation}
\dot{n}_m = -3 H n_m + \frac{11 m^5 T}{128 \pi^3 \bar{M}_P^2} K_1(m/T),
\end{equation}
where $K_1(x)$ is a modified Bessel function of the second kind and
we have assumed that $m_e = 0$. This assumption has very little
influence on the results.

\subsection{Production during the radiation dominated epoch}
The universe enters the radiation dominated epoch at some
temperature $T$, which we shall refer to as the reheating
temperature, $T_{\rm RH}$. 
Production of KK modes during this epoch was studied in detail
by Hall and Smith \cite{hs99}, and in this section their results
are rederived.
The present
day number density can be found be integrating the Boltzmann
equation
\begin{equation}
n_0(m) \simeq \frac{19}{64 \pi^3} g_{*,RH}^{-1/2} T_0^3
\frac{m}{\bar{M}_{P}} e^{-\Gamma_{{\rm decay},m}t_0}
\int_{m/T_{RH}}^\infty q^3 K_1(q) dq.
\end{equation}
This equation applies to the number density for one mode with
mass $m$. However, if we are interested in the total present
day contribution to the mass density from all modes, then
we need to sum over all modes. This sum can be replaced by an
integral over $dm$ because the mode density is very high
\cite{hanhart}.
This integration yields the result
\begin{eqnarray}
\rho_{0,{\rm thermal}} & \simeq & 1.9 \times 10^{-22} S_{n-1} {\rm GeV}^4 
\left(\frac{T_{RH}}{M}\right)^{n+2} \nonumber \\
&& \,\, \times \int_0^\infty dz \, z^{n+1}
e^{-\Gamma_{{\rm decay},m}t_0} \int_z^\infty dq q^3 K_1(q),
\label{eq:thermal}
\end{eqnarray}
where $S_{n-1} = 2 \pi^{n/2}/\Gamma(n/2)$.
In Fig.~1 we show the constraints on $M$ from demanding that
$\rho_0 \leq \rho_{\rm crit}$, for the case of $n=2$.
This result is identical to what was found in Ref.~\cite{hs99}.
\begin{figure}[h]
\begin{center}
\epsfysize=7truecm\epsfbox{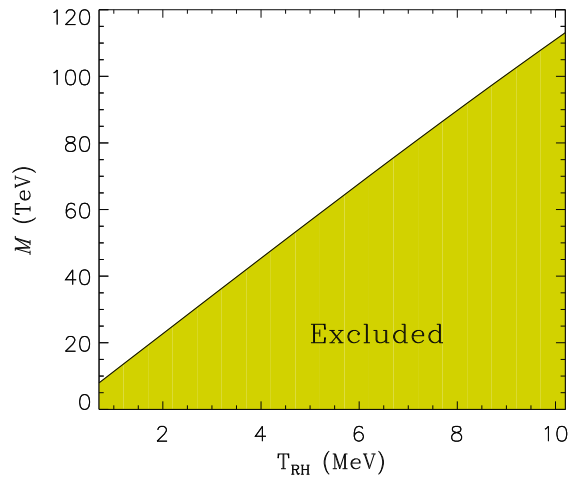}
\vspace{0truecm}
\end{center}
\caption{The lower bound on $M$ as a function of $T_{RH}$, when only
modes produced during the radiation dominated epoch are considered.
The calculations assume that $h = 0.75$ and $n=2$.}
\label{fig1}
\end{figure}

From this it is evident that for $T_{RH} \gtrsim 3$ 
MeV the bound is tighter
than what is found from SN1987A. 
However, it was shown in Refs.~\cite{kks99,kks00} that
$T_{RH} = 0.7$ MeV can still be compatible with BBN, so if $T_{RH}$
is sufficiently low the overproduction of KK states can be avoided.

\subsection{Production during reheating}

In the above treatment it was assumed that the universe enters
the radiation dominated epoch instantaneously at the reheating
temperature. However, this is not the case for any physically
acceptable scenario. Plausibly, the universe enters 
the radiation epoch after some reheating by the decay of
a massive scalar field (or by some other means of entropy
production). 
The only reasonable alternative is that the radiation dominated
epoch extended to much higher temperatures (of the order $M$).
Here, we look at the ``standard'' case where
reheating occurs from the decay of the inflaton field
(for further discussion of inflation in scenarios with large 
extra dimensions, see for instance Ref.~\cite{cgt99}
and references therein).

What happens is that the universe starts reheating when the
inflaton enters the oscillating regime. The important parameters
are the density, $\rho_{\phi,i}$, of the inflaton when reheating 
begins and the decay rate of the inflaton, $\Gamma_\phi$.
The Boltzmann equations for this system have been solved numerous
times (see e.g.\ Refs.\ \cite{kolb,ckr99}). 
The result is that the temperature of the produced radiation
immediately increases to a maximum value which depends on 
$\rho_{\phi,i}$. After this, there is a period of continual entropy
production during which the universe is matter
dominated by the $\phi$ field and $T \propto t^{-1/4}$ (as opposed
to the case where no entropy is produced, $T \propto t^{-1/2}$).
At the time $t \simeq \Gamma_\phi^{-1}$ the inflaton decays
rapidly and the universe becomes radiation dominated.
Using this, it is easy to calculate the number of KK modes
produced during the reheating phase. 
$\Gamma_\phi$ is directly related to $T_{RH}$ by $T_{RH}
\simeq 0.5 \sqrt{\Gamma_\phi \bar{M}_P}$ \cite{kks00}, but the
additional parameter $\rho_{\phi,i}$ is introduced in the
analysis.
However, instead of this we use the parameter
$\alpha \equiv T_{MAX}/T_{RH}$, where $T_{MAX}$ is the maximum
temperature reached during reheating (the relation between
$\rho_{\phi,i}$ and $T_{MAX}$ is given in Ref.~\cite{gkr00}).
From this, one gets an expression completely equivalent to 
Eq.~(\ref{eq:thermal}).
During reheating entropy is produced continuously. The entropy
density is given by $s = g_* T^3$ at all times, where $g_*$ is
the number of relativistic degrees of freedom contributing to the
entropy \cite{kolb}. We assume that $g_* \simeq 10.75$ at all times.
Although this is not the case at very high temperatures, the assumption
introduces only a modest error.
Since $T \propto t^{-1/4}$ and $a \propto t^{2/3}$
(because the universe is matter dominated), the number density of a
KK-mode (if one ignores production) is $n_{m} \propto T^{8}$.
This yields $n_{m}/s \propto T^{5}$ during reheating.
In order to solve the Boltzmann equation during reheating we
introduce the variable $X_m \equiv n_{m} s^{-1} (T/T_{RH})^{-5}$ which is
constant during reheating if there is no production.
Then the Boltzmann equation can be simply written as
\begin{equation}
\dot{X}_m = \frac{1}{s (T/T_{RH})^5} \frac{m^5 T}
{128 \pi^3 \bar{M}_P^2} K_1(m/T).
\end{equation}
By integration this gives
\begin{equation}
X_{m,RH} = \frac{4}{128 \pi^3  \bar{M}_P^2} 
\frac{t_{RH} T_{RH}^{13}}{m^6 s_{RH}} \int_{m/T_{MAX}}^{m/T_{RH}}
dq q^{10} K_1(q),
\end{equation}
where we can approximate $t_{RH}$ by $t_{RH} \simeq 1.5 g_*^{-1/2} \bar{M}_P
T_{RH}^{-2}$ \cite{hs99}. The present day number density is then given
by $n_{m,0} = X_{m,RH} s_0$. The total density of all KK-modes can be found
be integrating over $dm$ \cite{hs99}, and gives
\begin{eqnarray}
\rho_{0,RH} & \simeq & 1.9 \times 10^{-22} S_{n-1} {\rm GeV}^4 
\left(\frac{T_{RH}}{M}\right)^{n+2} \nonumber \\
&& \,\, \times \int_0^\infty 2 dz \, z^{n-6} 
e^{-\Gamma_{{\rm decay},m}t_0}\int_{z/\alpha}^z dq q^{10} K_1(q).
\label{eq:reheat}
\end{eqnarray}
The factor $e^{-\Gamma_{{\rm decay},m}t_0}$ comes from the fact that
some of the produced KK-modes decay before the present.
This is one of the main new result of the present paper.
Notice that Eq.~(\ref{eq:reheat}) is very similar to Eq.~(\ref{eq:thermal}),
except for the fact that $z^{n+1} q^3$ is changed to $z^{n-6} q^{10}$. This 
can be understood quite simply. Modes produced during reheating are
diluted by entropy production relative to modes produced during the
thermal epoch by the factor $(T/T_{RH})^{-5}$. Furthermore, the time-interval
for production during reheating is different from what it is during
the thermal epoch. During reheating $dt/dT \propto T^{-5}$, whereas
during the thermal epoch $dt/dT \propto T^{-3}$. Altogether this gives
a factor $(T/T_{RH})^{-7} = z^{-7} q^{7}$ difference, which is what is found.
\begin{figure}[h]
\begin{center}
\vspace*{-0.5truecm}
\epsfysize=7truecm\epsfbox{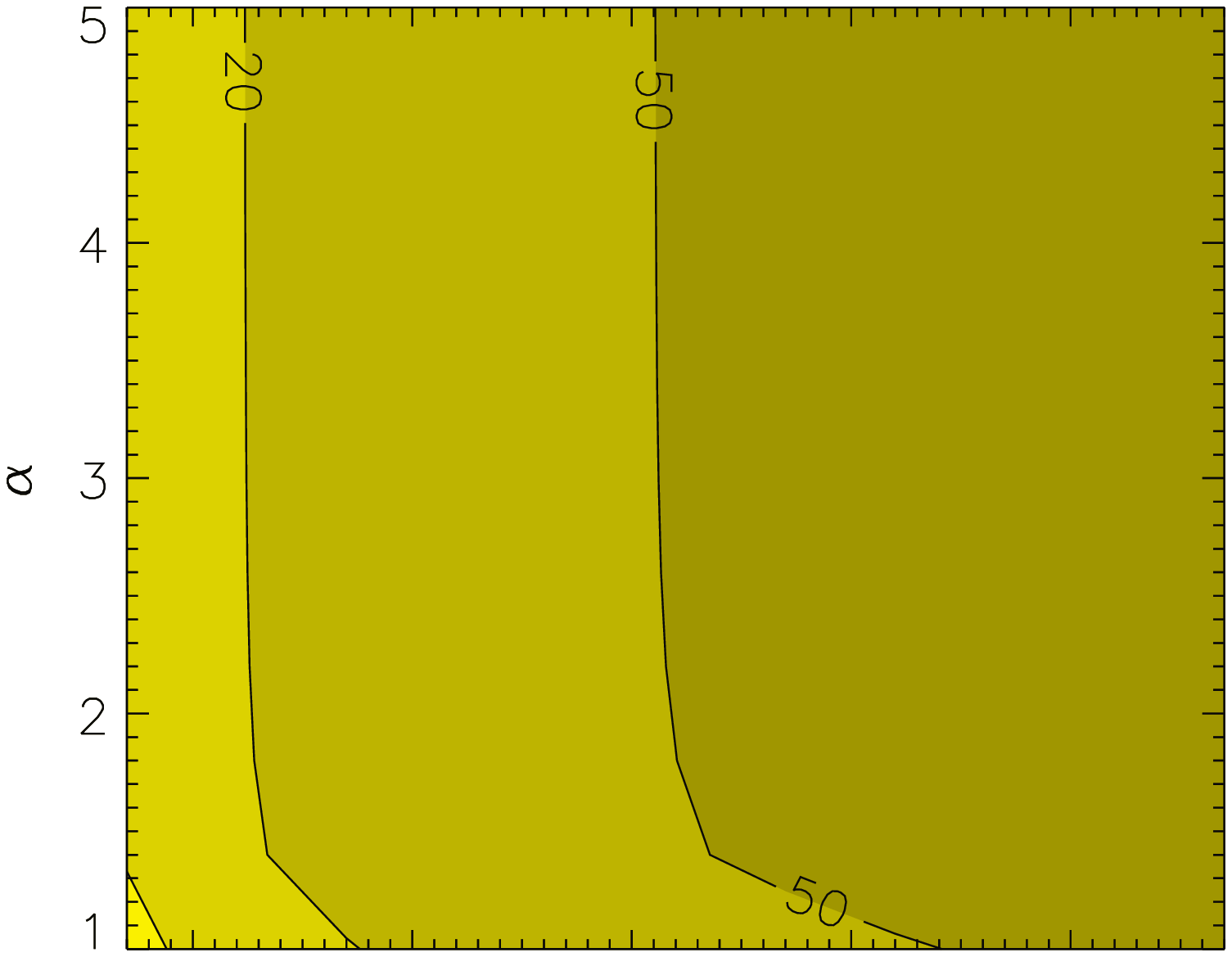}
\vspace*{-1truecm}
\epsfysize=7truecm\epsfbox{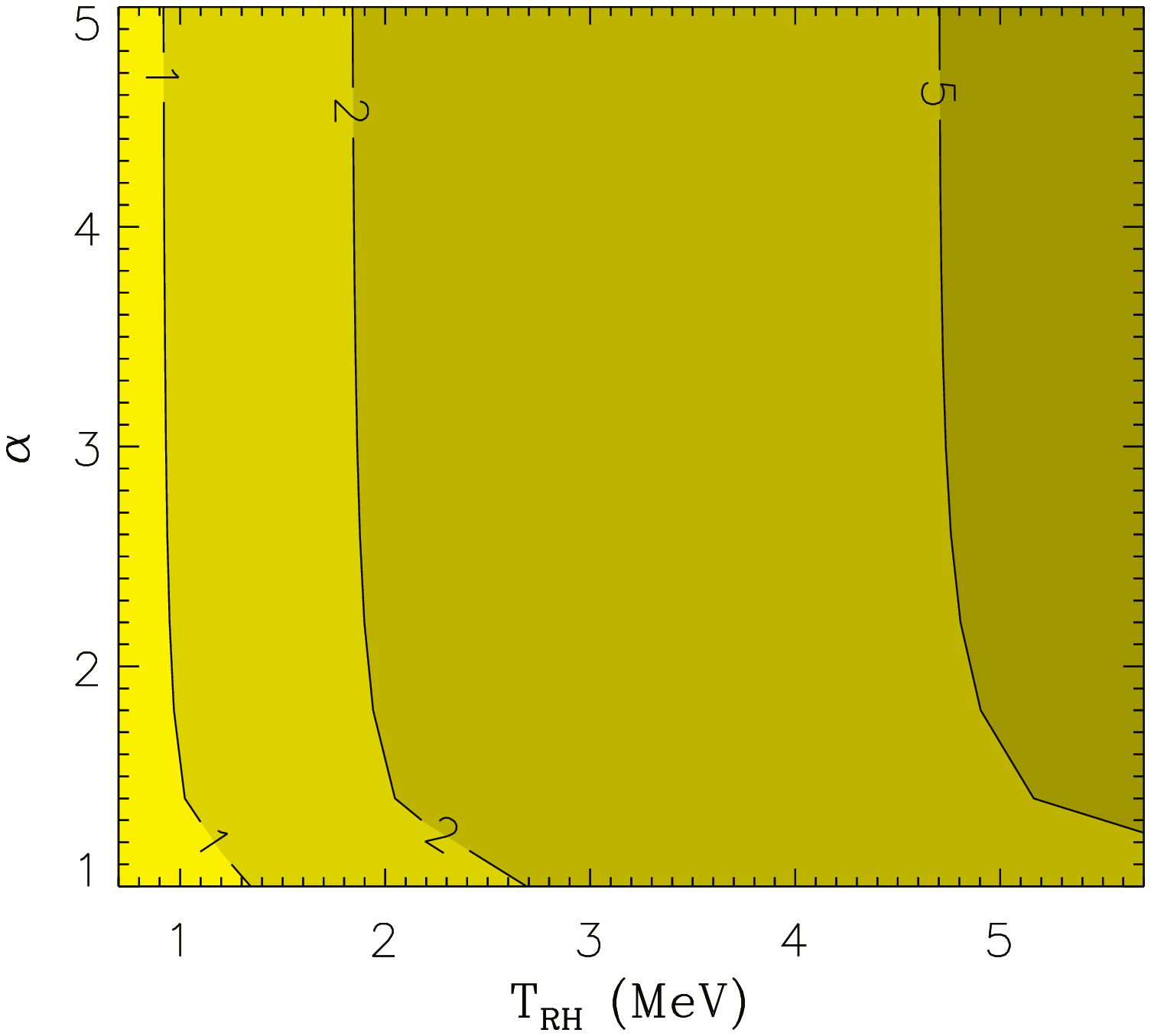}
\vspace{0truecm}
\end{center}
\caption{The lower bound on $M/{\rm TeV}$ as a function of
$T_{RH}$ and $\alpha$, from demanding that $\rho_{0,{\rm thermal}}
+\rho_{0,RH} \leq
\rho_{\rm crit}$. 
The upper panel is for $n=2$ and the lower for $n=3$.
The value $h=0.75$ has been used.}
\label{fig2}
\end{figure}

Fig.~2 shows the lower bound on $M$ for $n=2$ and 3,
as a function of 
$T_{RH}$ and $\alpha$.
It is evident from the figures that the contours quickly become
vertical for increasing $\alpha$. This happens because the 
KK modes produced at early times during reheating have been
diluted away by entropy production so that they do not contribute
significantly at late times.
Even so, accounting for the modes produced during reheating does
strengthen the bound on $M$. For the case of $n=2$ the bound 
is 6.8 TeV if only the thermal production is included, whereas it is
9.7 TeV if reheating is also accounted for
(assuming $T_{MAX} \gg T_{RH}$). For $n=3$ these figures
are 0.44 TeV and 0.66 TeV respectively.

Notice that this result is different from the result found
by Giudice et al.~\cite{gkr00}, that the final abundance depends only
on $T_{RH}$, and not on $T_{MAX}$. 
The reason is that there is a dense spectrum of modes with 
different masses. The abundance of each mode obeys the relations found
in Ref.~\cite{gkr00}, i.e.\ that the final abundance for any
mode with $m \lesssim {\rm few} \times T_{MAX}$ is independent
of $T_{MAX}$. However, for higher $T_{MAX}$, many more modes are
excited, and the end result is that the total production of KK
modes increases with increasing $T_{MAX}$.

Nevertheless, if we only consider the bound on $M$ from the demanding
that $\rho_0 < \rho_{\rm crit}$, it is significantly weaker than
the bound from SN1987a. In the next section we consider the much
stronger bound from the diffuse gamma background.

\section{Constraints from the diffuse gamma background}
Apart from the production mechanisms there is also the possibility that 
the massive KK states decay into particles on the brane.
The decay rate for different branches has been calculated by
Han, Lykken and Zhang \cite{hlz99}.
The decay rate into particles on the brane is of the order
$\Gamma \sim m^3/M_P^2$ for any kinematically allowed final state.
With the temperatures discussed here, the decay lifetime
is on most cases of the same order of magnitude as the Hubble time.
This means that there will be visible effects,
especially from
the decay contribution to the diffuse gamma background.
It was shown by Hall and Smith \cite{hs99}
(see also Ref.~\cite{bd99}) that this leads to a very
stringent constraint on $M$, even with $T_{RH} = 1$ MeV. The modes
produced during reheating have higher mass and therefore much higher 
decay rates. This means that even tighter constraints can be put
on $M$ if reheating is included.

For $n=2$, the contribution to the diffuse gamma background from KK decays
is given by
\begin{eqnarray}
\frac{dn}{dE} & \simeq & 345 M_{\rm TeV}^{-4} T_{\rm RH, MeV}^{5}
\left(\frac{t_0}{10^{10} y}\right) {\rm MeV}^{-1} {\rm cm}^{-2}
{\rm s}^{-1} {\rm ster}^{-1} \nonumber \\
&& \times \beta^{1/2}
\left[ \int_{2 \beta}^{z_{\rm max}} dz z^{7/2} E(z)
\int_z^\infty q^3 K_1(q) dq + \right. \nonumber \\
&& \,\,\, \left. \int_{2 \beta}^{z_{\rm max}} 2 dz z^{-7/2}E(z) 
\int_{z/\alpha}^z q^{10} K_1(q) dq \right],
\label{eq:diffuse}
\end{eqnarray}
where $\beta = E/T_{RH}$ and 
$E(z) = \exp(-3.3 \times 10^{-7} z^{3/2} T_{RH,MeV}^3 
t_0(10^{10}y) \beta^{3/2})$.
As the upper limit for the mass integral, $z_{\rm max}$, we take
$z_{\rm max} = 2.7 \times 10^{3} T_{RH,MeV}$, because KK modes
above this mass decay before CMBR formation, and therefore do not
contribute to the present diffuse flux.
Note that for $\alpha=1$ this equation is equivalent to what
is found in Ref.~\cite{hs99}. Also note that there is the same factor
$q^7 z^{-7}$ difference between the thermal (the first part in the bracket)
and the reheating (the second part) contributions as was found between
Eqs.~(\ref{eq:thermal}) and (\ref{eq:reheat}).
For higher $n$ one obtains expressions very similar to Eq.~(\ref{eq:diffuse}).
Observationally, the diffuse gamma background in the MeV range has been
measured by the EGRET \cite{egret,rothstein}
($30-10^4$ MeV) and COMPTEL \cite{comptel,rothstein} (0.8 - 30 MeV) 
experiments. The flux measured by EGRET is approximately
$\frac{dn}{dE} = 2.3 \times 10^{-3} (E/{\rm MeV})^{-2.07}
{\rm MeV}^{-1} {\rm cm}^{-2}
{\rm s}^{-1} {\rm ster}^{-1}$, and the flux measured by COMPTEL is
$\frac{dn}{dE} = 6.4 \times 10^{-3} (E/{\rm MeV})^{-2.3}
{\rm MeV}^{-1} {\rm cm}^{-2}
{\rm s}^{-1} {\rm ster}^{-1}$.
Demanding that $\frac{dn}{dE}_{KK} \leq \frac{dn}{dE}_{\rm obs}$
translates into the lower bound on $M$
shown in Fig.~3 as a function of $T_{RH}$ and $\alpha$.
Note that for the relatively low masses we study here, constraints
from light nuclei abundances \cite{jedamzik} are not very important.

\begin{figure}[h]
\begin{center}
\vspace*{-0.5truecm}
\epsfysize=7truecm\epsfbox{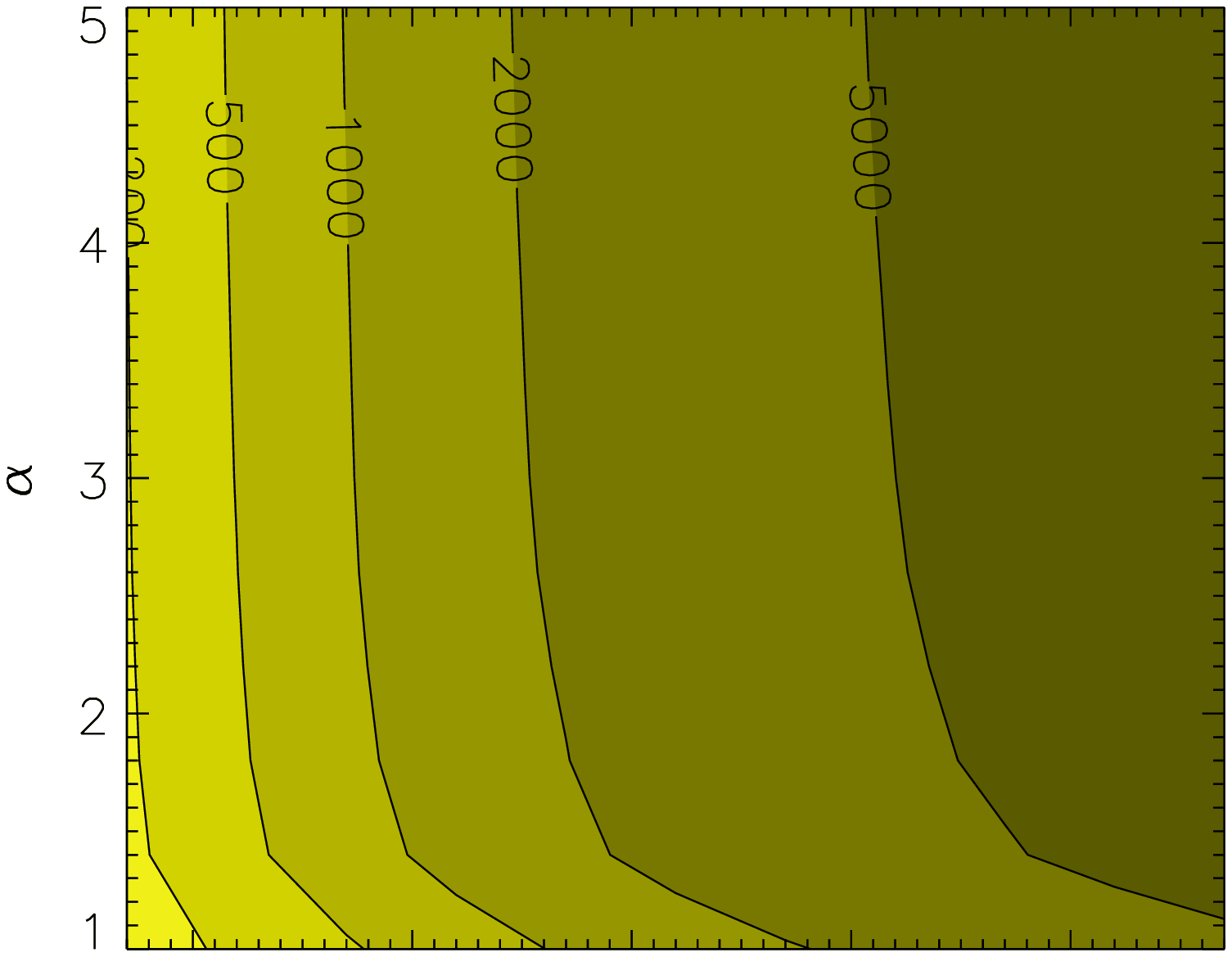}
\vspace*{-1truecm}
\epsfysize=7truecm\epsfbox{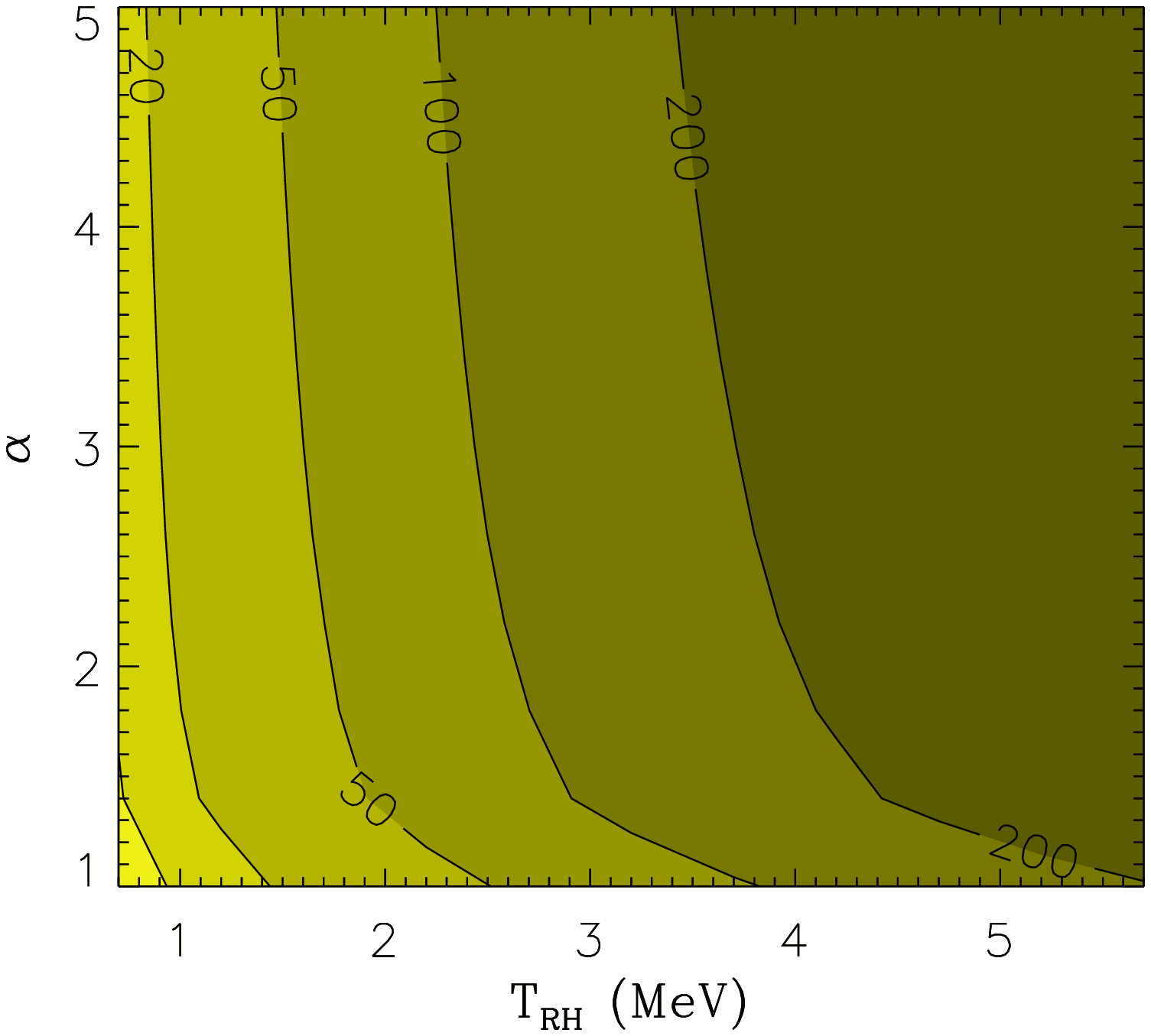}
\vspace{0truecm}
\end{center}
\caption{The lower bound on $M/{\rm TeV}$ as a function of
$T_{RH}$ and $\alpha$, from the diffuse gamma background. 
The upper panel is for $n=2$ and the lower for $n=3$.
The value $t_0 = 10^{10}y$ has been used.}
\label{fig3}
\end{figure}

From this figure it is evident that increasing $\alpha$ leads to
significantly stronger bounds on $M$. If $T_{RH} = 0.7$ MeV, then
for $n=2$ the bound goes from $M > 73$ TeV at $\alpha = 1$ to
$M > 167$ TeV at $\alpha = 1400$ ($T_{MAX} = 1$ GeV).
For $n=3$ the corresponding numbers are 
$M > 3.9$ TeV at $\alpha = 1$ to
$M > 21.7$ TeV at $\alpha = 1400$, a factor of 5.6 increase.
In most reasonable models it is difficult to obtain values
of $T_{MAX}$ which are smaller than 1 GeV, rather $T_{MAX}$ will
usually be much higher than 1 GeV.

Finally, it should be noted that as 
$n$ is increased, the difference between including reheating
and only treating thermally produced modes increases. 
In Table I the lower bound on $M$ is shown for different values
of $n$ and $T_{MAX}$ to illustrate this.

\narrowtext
\begin{table}
\caption{The lower bound on $M$ in TeV for different values of
$n$ and $T_{MAX}$. All values are for $T_{RH} = 0.7$ MeV.}
\begin{tabular}{lcccc}
n & $T_{MAX} = 0.7$ MeV & 50 MeV & 100 MeV & 1 GeV \\
\tableline 
2 & 73 & 161 & 165 & 167 \cr
3 & 3.9 & 16.0 & 18.6 & 21.7 \cr
4 & 0.47 & 2.96 & 3.75 & 4.75 \cr
5 & 0.10 & 0.89 & 1.19 & 1.55
\end{tabular}
\end{table}

For masses which are low enough that the KK modes have not
decayed away before the present, the lower bound on $M$ as a function
of $T_{RH}$ and $T_{MAX}$ exhibits a quite simple behaviour.
For $\alpha=1$, $M_{\rm min} \propto T_{RH}^{(n+5)/(n+2)}$, whereas
for constant $T_{RH}$ and large $\alpha$, 
$M_{\rm min} \propto T_{MAX}^{(n-2)/(n+2)}$.
This means that for $n \geq 3$ the contribution from modes produced
during reheating keeps increasing with increasing $T_{MAX}$.
However, for very high masses the modes have decayed away early on
so that $M_{\rm min}$ reaches a limiting value for high $T_{MAX}$.

\section{Discussion}
We have discussed in detail how KK modes are produced in the 
early universe. 
First, production during the radiation dominated epoch 
was discussed, and the results found in Ref.~\cite{hs99}
were rederived.

It was then shown that if reheating before the thermal epoch is
taken into account, the bound on the fundamental Planck scale, $M$,
is strengthened significantly and becomes stronger than the
supernova bound, albeit the theoretical uncertainty is larger.
This is the main result of the present paper.

We showed that if $T_{MAX}$ during reheating is 1 GeV, then the
bound on $M$ is $M>167$ TeV for $n=2$ and $M>21.7$ TeV for $n=3$.
These bound can be translated into upper bounds on the radii of the 
extra dimensions using the relation \cite{hs99}
\begin{equation}
R_{mm} = 2 \times 10^{31/n-16} \left(\frac{\rm 1 TeV}{M}\right)^{1+2/n}.
\end{equation}
From this, we get
$R< 2.2 \times 10^{-5}$ mm for $n=2$ and
$R< 2.5 \times 10^{-8}$ mm for $n=3$.

Note that in the present treatment we have assumed that the inflaton
only decays to matter on the brane. If gravitons are also produced
at reheating, the bounds are tightened.

We finish by discussing briefly the
few possibilities for avoiding the very stringent bound
obtained above. In our analysis we assumed a toroidal geometry for
the extra dimensions. However, 
other choices of geometry lead to different spectra of KK modes.
As shown in Ref.~\cite{kaloper2}, compact hyperbolic manifolds
can lead to spectra with a lightest mode with $m \gtrsim 10$ GeV
and large energy spacing. In this case the cosmological bounds disappear
completely (as does all other astrophysical bounds).
The second possibility is that there are more branes embedded in the bulk.
In that case, it is possible that the KK modes have fast decay channels
to light particles on other branes. This would mean that the 
KK modes could have decayed into radiation before the present
\cite{add99}.

\acknowledgements
I wish to thank Sacha Davidson for pointing
out a serious error in the original version of the manuscript.

\end{document}